# A Fully Screen-Printed Vanadium-Dioxide Switches Based Wideband Reconfigurable Intelligent Surface for 5G Bands

Yiming Yang, *Student Member, IEEE,* Mohammad Vaseem, *Member, IEEE,* Ruiqi Wang, Behrooz Makki, *Senior Member, IEEE*, and Atif Shamim, *Fellow, IEEE*

*Abstract*— Reconfigurable Intelligent Surface (RIS) is attracting more and more research interest because of its ability to reprogram the radio environment. Designing and implementing the RIS, however, is challenging because of limitations of printed circuit board (PCB) technology related to manufacturing of large sizes as well as the cost of switches. Thus, a low-cost manufacturing process suitable for large size and volume of devices, such as screen-printing is necessary. In this paper, for the first time, a fully screen-printed reconfigurable intelligent surface (RIS) with vanadium dioxide ($VO_2$) switches for 5G and beyond communications is proposed. A $VO_2$ ink has been prepared and batches of switches have been printed and integrated with the resonator elements. These switches are a fraction of the cost of commercial switches. Furthermore, the printing of these switches directly on metal patterns negates the need of any minute soldering of the switches. To avoid the complications of multilayer printing and realizing the RIS without vias, the resonators and the biasing lines are realized on a single layer. However, this introduces the challenge of interference between the biasing lines and the resonators, which is tackled in this work by designing the bias lines as part of the resonator. By adjusting the unit cell periodicity and the dimension of the H-shaped resonator, we achieve a 220° to 170° phase shift from 23.5 GHz to 29.5 GHz covering both n257 and n258 bands. Inside the wide bandwidth, the maximum ON reflection magnitude is 74%, and the maximum OFF magnitude is 94%. The RIS array comprises 20 × 20 unit cells ($4.54 \times 4.54\lambda^2$ at 29.5 GHz). Each column of unit cells is serially connected to a current biasing circuit. To validate the array's performance, we conduct full-wave simulations as well as near-field and far-field measurements. The fully printed array shows signal enhancements of around 8-10 dB, validating its effectiveness.

## I. INTRODUCTION

WITH the increasing demand for higher capacity and lower latency of 5G and beyond communications, the carrier frequency of electromagnetic (EM) wave increases from below 5 GHz to more than 20 GHz. As the wavelength decreases from meters to millimeters, the EM wave experiences a higher loss from the atmosphere and can be more easily blocked by obstacles. To solve this issue, direct line-of-sight (LOS) communication is preferred and beam-forming technologies are adopted for the transmitter/receiver antennas and circuits, which increases design complexity, cost, and, power consumption [1]. Reconfigurable Intelligent Surfaces (RIS), sometimes referred to as intelligent reflecting surfaces, offer a more efficient solution by adding reconfigurability to the environment of EM wave propagation [2]. RIS consists of periodic elements, and their reflective/transmissive phase and magnitude are reconfigurable based on the requirement of the RIS surface impedance profile so that it can redirect and focus the reflection/transmission wave to the devices. Thus, the RIS installed in the environment can improve communication quality by helping establish a secondary LOS between the base stations and the users.

RIS is currently being investigated in one of the European telecommunications standards institute (ETSI) industry specification groups [3]. In the 3rd generation partnership Project (3GPP), RIS has been proposed as a discussion topic in both Releases 18 and 19 (not as a specification yet), and is believed as a 6G technology that can assist the base stations in blind spot removal, especially at higher frequency bands [4]. There have been many theoretical works to characterize and optimize the RIS-assisted communication channels [5 - 7], but the implementation for 5G and beyond RIS is still limited and thus gaining more and more research interests. As the number of elements increases, the RIS has more reconfigurable surface area, more degree of freedom, and better performance, at the cost of additional complexity [8], [9]. Thus, a practical application of RIS would require hundreds or thousands of elements. For example, in [10], the mmWave RIS has 20×20 elements. Traditionally, RIS is fabricated with printed circuit

This work is supported by Ericsson Research under Grant OSR#4606.

Yiming Yang, Ruiqi Wang, Mohammad Vaseem, and Atif Shamim are with Computer, Electrical and Mathematical Sciences and Engineering Division, King Abdullah University of Science and Technology (KAUST), Thuwal, 23955-6900, Saudi Arabia (email: yiming.yang@kaust.edu.sa, ruiqi.wang.1@kaust.edu.sa, mohammad.vaseem@kaust.edu.sa, atif.shamim@kaust.edu.sa).

Behrooz Makki is with Ericsson Research, Ericsson, 417 56 Gothenburg, Sweden (email: behrooz.makki@ericsson.com).

boards (PCB) and the reconfigurability of unit cells is dependent on surface-mounted devices such as PIN diodes, MEMS switches, varactors, etc. The typical cost of one PIN diode, i.e., MACOM MA4AGBLP912 [11], is 7.5 USD. Even for a small RIS (20 x 20) with 400 elements, only the cost of switches would be 3000 USD. Meanwhile, as the array size increases, the fabrication cost increases significantly due to the subtractive manufacturing of PCBs and component mounting processes. To reduce the cost and enhance the speed of manufacturing, additive manufacturing techniques, such as screen-printing, are attractive options. This is particularly important because one of the main motivations for RIS is cost reduction, otherwise an expensive RIS will lose in competition with alternative expensive technologies such as network-controlled repeaters [12] and integrated access and backhaul networks [13], in terms of performance.

Though screen printing of conductors has become feasible in recent years [14 -18], printing of reconfigurable components, such as switches, is not common, particularly for higher frequency bands. Our group has prepared a screen printable $VO_2$ ink [19] and demonstrated the fully screen-printed $VO_2$ based Radio frequency (RF) switches which were characterized up to 20 GHz [19]. To achieve an acceptable insertion loss, multiple layers (up to 8 layers) of $VO_2$ were printed in [19]. Nonetheless, to fully print the RIS (with $VO_2$ switches also printed), there are several challenges that need to be solved. First, at millimeter wave frequencies, the unit cell requires a lower switch-on resistance, as compared to lower frequencies [19 – 21]. This requires better ink preparation process, which includes preparation of higher-quality $VO_2$ micro-particles, and better ink viscosity control. During printing, both the number of printed layers as well as the ink drying time and temperature should be carefully chosen so that the switches have good ON resistances and ON/OFF ratios simultaneously. Second, a uniform, and repeatable switch performance is required for the complete batch of $VO_2$ switch array. This requires special stencil design and multi-layer alignment process.

On the other hand, to make the resonator design compatible with the batch $VO_2$ printing process, a single-layer, via-less resonator design is preferred, in contrast to the traditional PCB designs with multiple layers and vias [22]. However, this results in coupling between the biasing lines and the resonators, which negatively influences the performance of the RIS. We have addressed the coupling issue by integrating the biasing lines as part of the resonators. In addition, designing the unit cell with a large bandwidth (23.5-29.5 GHz) is quite challenging. As frequency increases, common methods such as multiple resonators [23] become difficult because of the wavelength reduction and limitation of manufacturing resolution. We have designed the wideband unit cell by limiting the periodicity below half wavelength.

In this work we resolve many challenges in order to fully print the RIS. These range from better ink preparation and optimized printing recipes for both the conductor and $VO_2$ inks. As compared to our previous work, we can get a decent switch performance with 4 layers of printed $VO_2$ ink, which is half of the number of layers in [21]. On the design side, we propose a single-layer resonator design that integrates both the resonator and biasing lines on the same layer along the two orthogonal polarization directions. The design does not need any vias, and is thus compatible with the screen-printing process. To cover the bandwidth from 23.5GHz to 29.5GHz, we reduce the unit cell periodicity [24] and tune the switch resistance to maintain a smooth phase transition inside the large bandwidth, while keeping the cost and fabrication complexity low. To bias and configure the RIS pattern, we design and fabricate a controller board with current feedback control. We have performed, both the near-field and far-field, testing of the RIS using bistatic measurement setups. Our results show that the fully-printed RIS provides 8-10 dB signal enhancement in 0° anomalous reflection direction compared with specular reflection with -30° incidence in the near-field setup and 7.5 dB enhancement in the far-field setup with a 45° angular stability.

II. $VO_2$ PRINTING AND CHARACTERIZATION

*A. Ink Preparation*

$VO_2$ as a phase change material is popular for many RF and optical applications. However, all of the previous works related to $VO_2$ have been done through high vacuum nano-fabrication processes in the clean room environment. As mentioned in the Introduction, our group has previously demonstrated fully screen-printed RF switches based on a custom $VO_2$ ink [21]. The details of the experimental procedure are described in [21]. In this work, we have adopted modified synthesis protocol to achieve higher crystallinity and purity of $VO_2$ micro-particles (MPs) for high performance switches with reduced number of printed layers. The $VO_2$ MPs have been prepared using high purity vanadium (V) oxide ($V_2O_5$) based precursor (99.96%). The crystallinity of as-prepared $VO_2$ MPs has been verified by X-ray diffraction (XRD) analysis. XRD spectra (Fig. 1(a)) confirms the high crystalline peaks corresponding to the $VO_2$ (M) phase without any impurities. The $VO_2$ MPs are further tested for differential scanning calorimetry (DSC) analysis to confirm the Metal-Insulator-Transition (MIT) behavior. Clear MIT peaks, at 67°C during heating (endothermic) and 56°C during cooling (exothermic), can be observed (Fig. 1(b)). The DSC analysis confirms the reversible phase transition characteristics from monoclinic to tetragonal structures. The monoclinic phase corresponds to an insulator characteristic; however, the tetragonal, or rutile, phase corresponds to a metallic characteristic.

Once the MIT behavior and crystallinity are confirmed, the next step is to prepare the ink. To prepare the screen-printable ink, we load the $VO_2$ MPs into a controlled polymer formulation. First, a viscous organic binder-based polymer solution is formulated by mixing ethyl cellulose (EC), terpineol, and ethanol at a weight ratio of 1:4:0.4. Then, the $VO_2$ MPs are mixed with the viscous polymer solution. As compared to our previous work [21], we have chosen a higher loading of $VO_2$ MPs at a weight ratio of 1:1. This step is followed by agitation to obtain the $VO_2$ ink. To ensure optimized printing, we control the printing speed and number of over-printed layers. The SEM image of the $VO_2$ ink (Fig. 1(c)) shows that the $VO_2$ MPs are evenly dispersed in the polymer ink and that the MPs size is below 5 μm.

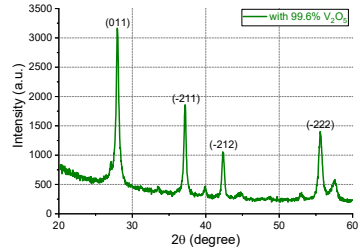

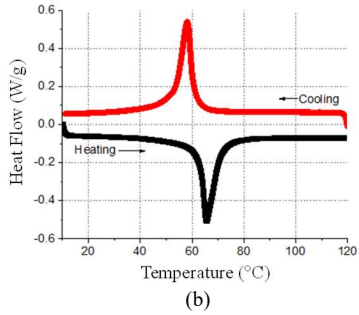

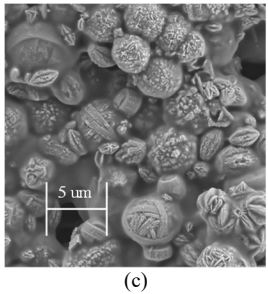

Fig. 1. Characterization of the $VO_2$ micro-particles. (a) XRD analysis of $VO_2$ (M) micro-particles. (b) DSC analysis of $VO_2$ micro-particles. (c) SEM analysis of $VO_2$ ink.

### B. Switch Printing

After the preparation of the ink, screen-printing recipe has been optimized for printing of the RF structures, as shown in Fig. 2. A professional screen-printing system (AUREL 900PA) has been utilized in this work. This printer has a 21 inch × 21inch stainless-steel screen mesh mask with 325 mesh count, 22.5° mesh angle, and 10 μm emulsion thickness. For uniform printing, an optimized speed of 150 mm s−1 has been selected. For the printing of conductor layers, an Ag paste has been utilized. With the Stencil Mask in Step 1 of Fig. 2(a), we can squeeze the Ag Paste through the conductor pattern, and thus electrodes are printed on the PEN substrate. After drying the Ag ink in an oven, we use another stencil with the switch pattern to print the $VO_2$ ink layer, as shown in Step 2 of Fig.2. With the cross markings, we can align and print the $VO_2$ ink on top of the Ag paste conductor, forming good contact between the conductor and switch devices.

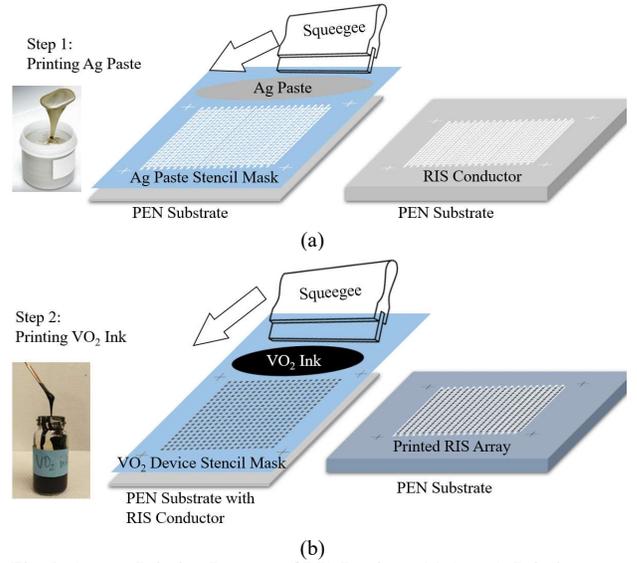

Fig. 2. Screen-Printing Process of RIS Devices. (a) Step 1: Printing Ag Paste. (b) Step 2: Printing $VO_2$ Switch.

### C. Switch Characterization

To characterize and compare the performance of screen-printed $VO_2$ switches and commercial PIN diodes, we fabricated the shunt switches on the 100um gap GSG lines as shown in Fig. 3. The GSG lines are 10mm long and we measure the isolation $S_{12}$ when the switch is ON, and the insertion loss when the switch is OFF with a probe station at 20 GHz and 40 GHz. To measure the $VO_2$ with different ON resistances, we printed $VO_2$ samples with 2, 4, and 6 layers to control their thickness with the same batch of $VO_2$ ink. The 2-probe measurement of the 4-layer switch shows an ON resistance of 3.5 ohm and an OFF resistance of 1 kohm, with an OFF/ON ratio of 285. The OFF-ON transition time can be as short as 0.4 us [21]. The details of the characterization results are shown in TABEL I.

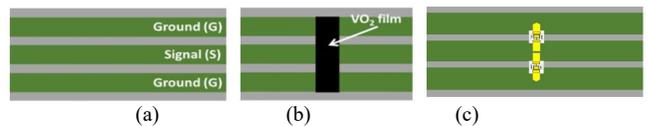

Fig. 3. Shunt switch characterization GSG lines. (a) reference. (b) screen-printed VO2 switch. (c) PIN diode MA4AGBLP912 switch.

TABLE I.
CHARACTERIZATION RESULT

| Sample | Shunt Switch ON | | Shunt Switch Off | |
|---|---|---|---|---|
| | Isolation (dB) @ 20GHz | Isolation (dB) @ 40GHz | Insertion Loss (dB) @ 20GHz | Insertion Loss (dB) @ 40GHz |
| Reference 100um Gap GSG Line | | | 0.65 | 0.59 |
| Shunt PIN Switch @ 20mA | **17.1** | **6.4** | **1.11** | **0.56** |
| 2 Layer VO$_2$ @ 200mA | 14.90 | 19.90 | 2.13 | 2.92 |
| 4 Layer VO$_2$ @ 200mA | **17.40** | **24.50** | **2.90** | **3.97** |
| 6 Layer VO$_2$ @ 200mA | 19.67 | 29.02 | 4.04 | 5.63 |

We can see that as the number of VO$_2$ printed layers increase, the isolation at the ON state increases due to a decrease in the ON resistance. The insertion loss at the OFF state also increases, which means that the OFF resistance also decreases. Compared with the PIN diodes, we can see that the 4-layer VO$_2$ has almost similar ON isolation at 20GHz. The 4-layer VO$_2$ switch has a similar ON resistance (4 ohm) as the PIN diode. From the datasheet, the PIN diode also has an additional 37fF shunt parasitic capacitance across the resistance and 1pH serial inductance. As frequency increases to 40GHz, the isolation of the PIN diode decreases, while the performance of VO$_2$ switches does not deteriorate. This shows that the VO$_2$ switches has a smaller parasitic inductance as compared to the commercial PIN diodes. The OFF resistance of the VO$_2$ material is 1 kohm. The PIN diode has an off resistance of 10 kohm. Thus, the insertion loss of VO$_2$ at the OFF state is larger than PIN diodes. To summarize, the 4-layer VO$_2$ switch has a similar ON resistance as the commercial PIN diodes, smaller parasitic parameters at above 20 GHz, and smaller OFF resistance than PIN diodes.

### III. VO$_2$ Unit Cell Design

The evolution of the unit cell design is explained in this section. In order to achieve a fully printed process, appropriate selection of conductors and substrates has been made in this work. Those chosen substrate is a 0.8 mm thick AF32 glass substrate (Dk = 5.1, tanδ = 0.0084), because the end application requires these RIS panels to be attached to the glass windows, etc. However, to ease the process of printing and sintering, a 50 um PEN film (Dk = 2.8, tanδ = 0.003) is utilized. The chosen conductor is a silver paste ($\sigma = 5 \times 10^6$ S/m, 0.2 um thick), which is suitable for our screen-printing process. The PEN film, after printing and sintering, is attached to the glass substrate, which is backed by a copper ground plane.

In order to incorporate the VO$_2$ switches in the EM simulation model, simply ON and OFF resistances have been obtained from the characterization performed in section II. An ON resistance of 4 ohm and an OFF resistance of 1 kohm has been obtained. Though the above mentioned values have been used in the EM simulations, a sensitivity analysis has been performed due to the variations in the printed switches thicknesses, where it is found that unit cell performance stays within the acceptable levels, provided the ON resistance stays between 4-10 ohms and the OFF resistance stays 1 kohm or more. The switch resistances have been incorporated in the EM simulation model as RLC boundary conditions.

For the unit cell design, a via-less structure is considered to simplify the printing process and also to avoid additional losses due to printed vias. The evolution of the design is shown in Fig. 4. As a starting point, a spiral cross dipole has been selected due to its inherent large bandwidth, as shown in step A of Fig. 4. The DC biasing lines are also on the top surface, along with the resonator. The second step (B) shows the optimization of biasing lines which have become a part of the resonator. The third step (C) focuses on reducing the number of switches in a unit cell. The final step (D) shows the change of biasing from parallel to serial mode. These evolution steps are described in more detail in the following sub-sections.

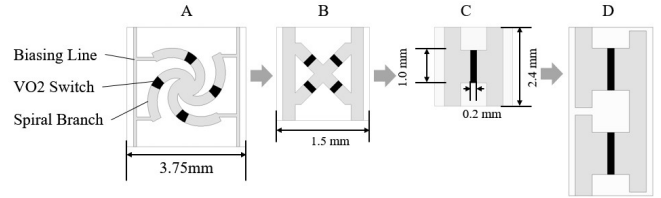

Fig. 4. Design Stages: (A) Resonator design with the bias lines (B) Optimizing the bias lines as part of resonator design, (C) Reducing the number of switches, (D) Adding slots for serial biasing.

#### A. Resonator design with the Bias Lines

The design of the screen-printable RIS unit cell begins with the spiral dipole design [25]. The simulations of the unit cell have been performed in HFSS with periodic boundary conditions. By turning the switches to ON and OFF conditions, the current path along the spiral branches experiences different lengths, providing a shift in the resonance frequency and phase shift in the desired bandwidth. As step A in Fig. 4 shows, a pair of biasing lines is connected to the edge of the resonator. The direction of the biasing lines is kept perpendicular to the polarization of the incident E field to minimize the effects of the biasing lines on the RF resonator performance. The phase of S$_{11}$ is the reflection phase of the RIS unit cell. In order to see the effects of biasing lines on resonator performance, simulations have been performed for the unit cell with and without the biasing lines, Fig.5 (a) and (b) illustrate the simulation results of the design in the ON state without and with the biasing lines, respectively. It can be observed that the resonant frequency moves down from 20 GHz to 9.8 GHz, and the resonance changes from normal resonance to anti-resonance due to increased loss of biasing lines. This means that the biasing lines have become part of the resonator. This can be confirmed by observing the current in the resonator at 9.8 GHz in Fig. 5 (c). We can see from the highlighted path that the RF current starts from one end of the biasing line, through the resonator, and stops at the other biasing line. So, the effective length of the resonator has increased because of the biasing lines. This implies that if we reduce the length of the resonator, the resonance current will have a smaller path, and thus a higher resonant frequency with a smaller loss.

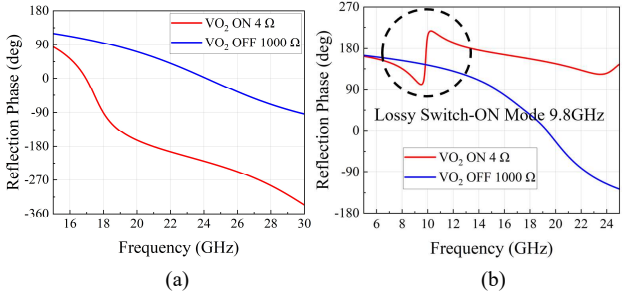
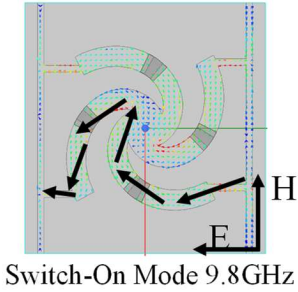

Fig. 5. Simulated Phase Shift of Unit Cells Before (a, b) and After (c, d) Adding Biasing Lines.

### B. Optimizing the the bias lines as part of Resonator Design

Comparing Step A in Fig. 4 and Fig. 6(a), the length of the spiral branch is reduced from 3.11 mm to 2.15 mm, and the simulation results are shown in Fig. 6(b). We can see that as the dipole length reduces, the resonant frequency increases from 9.8 GHz to 11 GHz and the response is recovered from anti-resonance to normal due to a smaller current length and conductor loss. By further tuning the dipole size, periodicity, and biasing line width we get the optimized unit cell design in Fig. 6(c). The optimized design has a reduced periodicity of 1.5mm. As the simulation result in Fig. 6(d) shows, the unit cell has a phase shift difference from 220 to 160 degrees in the ON and OFF states for the band between 22GHz to 30GHz. The small periodicity, however, requires a higher printing resolution. The dipole width is 0.2mm and the switch length is 0.1mm. Considering the resolution of screen printing, which is 150 ~ 200 um, printing this device could be challenging.

### C. Reducing the Number of Switches

It can be observed from Fig. 5(c) that the current distribution across the upper two and lower two branches is symmetric. Thus, the four branches of the cross dipole can be merged, as shown in Fig. 7(a). This not only solves the challenging printing issue mentioned previously, but also reduces the number of switches from four to one. In this version of the design, the $VO_2$ switch has a larger area and the conductor width is increased from 0.2 mm to 1 mm. Though, the switch length has been increased to 0.2 mm, it is still very close to the screen printing resolution, however, this issue can be resolved by creating additional contact regions between the $VO_2$ printed layer and the conductor to have better contact and printing quality. The simulated reflection phase and magnitude of the ON and OFF unit cells are shown in Fig. 7 (b) and (c).

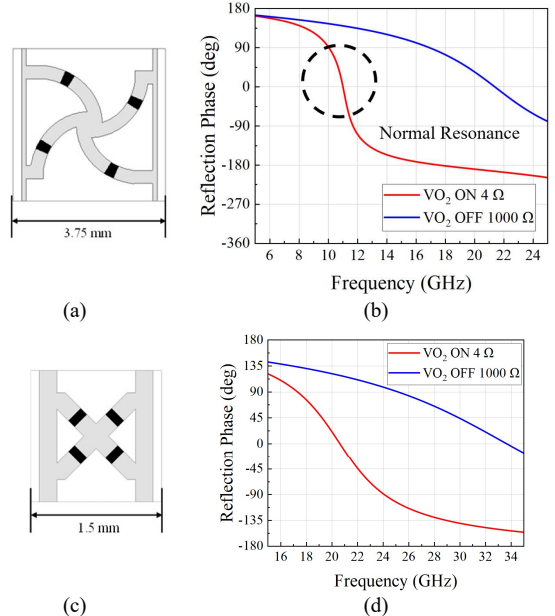

Fig. 6. Simulations of Unit Cells with (a, b) Reduced Dipole Length, and (c, d) Minimized Dipole Size

The reflection magnitude in the OFF state is above 87% across the bandwidth, and the reflection magnitude in the ON state ranges between 50% and 74% from 22.5GHz to 30GHz. The phase difference between the OFF and ON unit cells is 224° to 160° between 22.5 GHz and 30 GHz, which is close to 180° and is suitable for single bit RIS unit cell design.

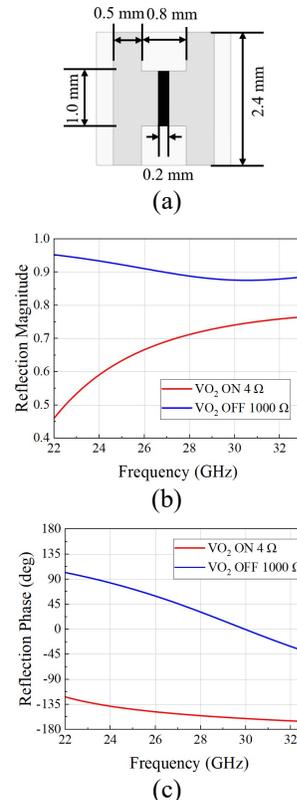

Fig. 7. (a) Optimized Unit Cell Design, (b) Simulated Magnitude, and (c) Phase Shift

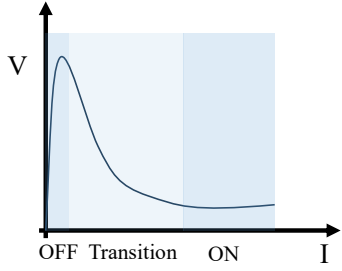

Fig. 8. I-V Characteristic of VO2 Material

### D. Adding Slots for Serial Biasing

If we build the array with the unit cell in Fig. 7(a), the unit cells along the same column are paralleled. This kind of biasing is suitable for voltage-controlled switches such as varactors. For $VO_2$ materials, however, parallel voltage biasing is not appropriate. This can be explained in the I-V curve of the $VO_2$ switch in Fig. 8. When the device is switched from OFF to ON state, we can see that in the transition region, the slope of voltage over current (AC resistance) is negative. When we apply constant voltage biasing, the device behavior is undetermined between OFF and Transition states and may even oscillate between them. However, if the current biasing is applied, the switch will always behave in a determined state. Thus, we convert the biasing from parallel to serial in the following steps.

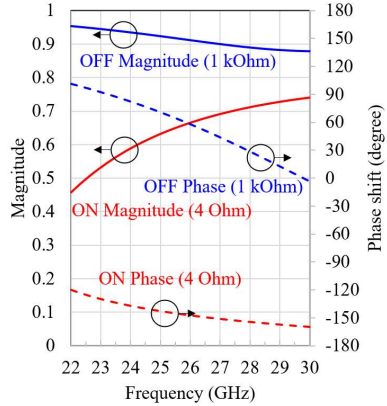

(a)

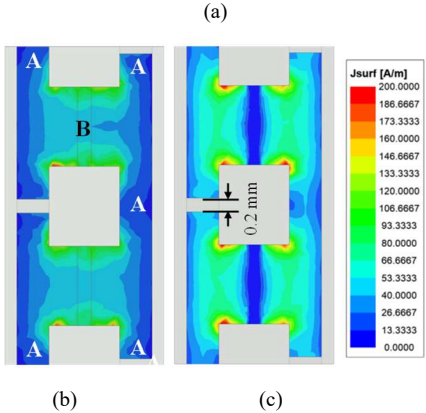

(b) (c)

Fig. 9. (a) Simulation Results and Current Distributions of the (b) ON and (c) OFF Unit Cells.

First, we study the current distribution shown in Fig. 5(c). We can see that at the edges of the biasing lines touching the unit cell boundary, there are current nulls. This implies that those areas have high impedance, and by adding 0.2mm cutouts in those areas, we can convert the resonator from parallel to serial connection, as shown in Fig. 9(c). By doing so, the periodicity is doubled, and the slots are interleaved from one dipole to another. The simulated magnitude and phase difference between the ON and OFF states are shown in Fig. 9(a), which are quite similar to the results shown in Fig. 7(a). The design has also been analyzed from the current distribution perspective and the results for 27.5GHz excitation are shown in in Fig. 9(b) and (c). It can be seen that the current has minimum magnitude at the edges as well as in the middle of the biasing lines (labeled with letter A in Fig. 9(b)). Further, the current has a maximum magnitude at the switch position (labeled with letter B in Fig. 9(b)). By turning the switch to ON and OFF conditions, the length of the current path can be controlled, and thus the resonant frequency can be shifted, which can provide the required phase shift difference between the states.

## IV. RIS ARRAY DESIGN AND SIMULATION

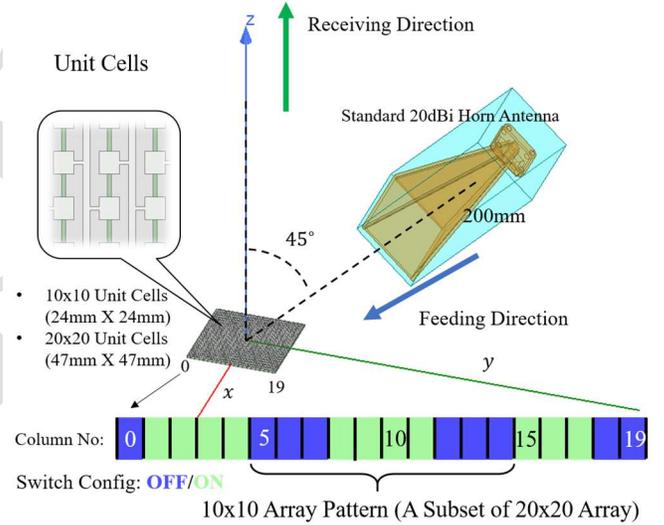

Fig. 10. Simulation model of the proposed RIS.

With the optimized unit cell for serial biasing, we construct and simulate the practical RIS model in ANSYS HFSS, as shown in Fig. 10. Here, two RIS configurations with 10 × 10 elements (10 reconfigurable columns) and 20 × 20 elements (20 reconfigurable columns) have been modeled. The practical sizes for these two RIS configurations are 24 mm × 24 mm and 47 mm × 47 mm, which corresponds to 2.2 $\lambda$ × 2.2 $\lambda$ and 4.3 $\lambda$ × 4.3 $\lambda$, respectively at the operation frequency of 27.5 GHz. A standard 20 dBi horn antenna has been used to illuminate the RIS from a distance of 200 mm. Considering the designed RIS unit cell has an angular stability of up to 45°, the horn antenna is placed at an incidence angle of 45°. It is believed that if the worst case scenario of 45° illumination provides acceptable results, other incident angles, such as 0°, 15°, and 30°, will demonstrate better performance according to the unit cell simulated results in Section III. Here, we assume the desired receiving direction to be 0°. Therefore, a specific RIS column

ON/OFF pattern is required for the 0° reflection, which can be obtained through the phase distribution design equation (1) [26].

$$\varphi_{i,j} = k|r^e_{i,j} - r^f| - k(u^r \cdot r^e_{i,j}) \quad (1)$$

In (1), $\varphi_{i,j}$ is the phase shift of the unit cell $(i,j)$, $k$ is the wave number at 27.5GHz, $r^e_{i,j}$ is the position of the unit cell $(i,j)$, $r^f$ is the location of the feed antenna, $u^r$ is the unit-length reflection direction vector.

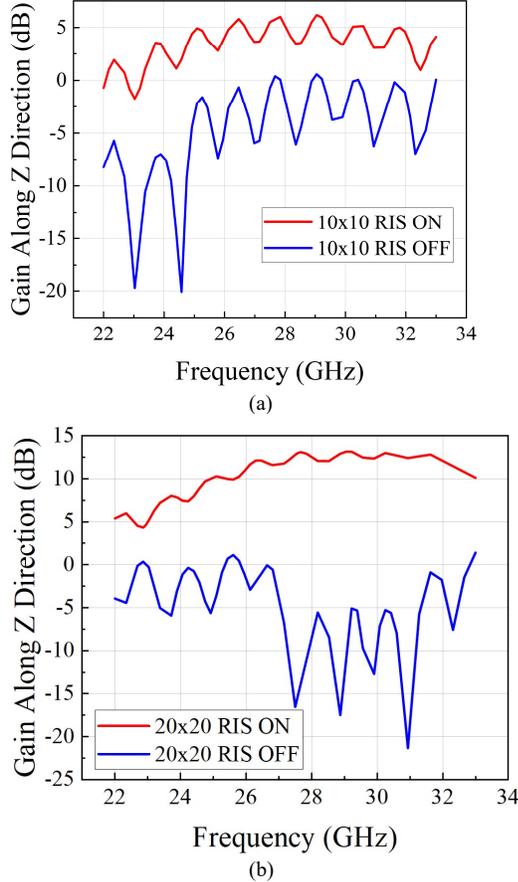

Fig. 11. Simulated gain enhancement of the proposed RIS when turning ON/OFF the RIS patterns.

The simulated results of the gain enhancement (the gain difference between the ON and OFF states) for the 10 × 10 and 20 × 20 configurations are shown in Fig. 11. It can be observed that from 23.5 GHz to 29.5 GHz, the RIS shows approximately 6 dB and 10 dB signal enhancement for the 10 × 10 and 20 × 20 arrays, respectively. The 3D radiation pattern of the proposed RIS with 20 × 20 configuration at 27.5 GHz is shown in Fig. 12. It can be seen that the RIS can help form a beam in the required angle (at around 0° reflection), compared with the null of the OFF state and the metallic surface.

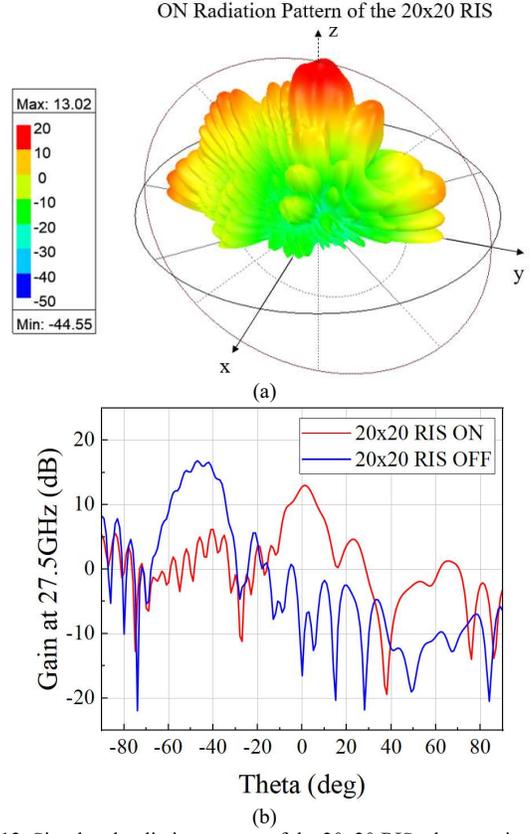

Fig. 12. Simulated radiation pattern of the 20x20 RIS when turning ON the RIS pattern at 27.5 GHz. (a) 3D plot. (b) 2D pattern in the y-z plane.

## V. ARRAY FABRICATION AND MEASUREMENT

### A. Fully-printed RIS Fabrication

The detailed fabrication process of the fully-printed RIS is shown as follows. We first follow the printing process shown in Fig. 2(a) to print the Ag paste on the PEN substrate. The quality of the printing can be observed in Fig. 13(a). The minimum gap is 0.2 mm, and it can be seen that the printed pattern matches well with the design. When the ink is dried, the $VO_2$ layer is printed on top of the Ag paste layer with the help of the cross-alignment marks at the corners of the array, as shown in Fig. 2(b). The $VO_2$ layer requires multiple batches of printing to increase the thickness. During each batch, the $VO_2$ layer is heated for several minutes with an infrared lamp to dry up the ink. This can prevent internal cracking of the switch after printing and drying. Finally, the $VO_2$-printed films are annealed at 210 $^0$C in vacuum. After printing many switches of varying heights and their 2-probe resistance characterization at 90 °C, it was concluded that the $VO_2$ switches should be thicker than six layers to get a switch-on resistance of less than 4 ohms. The fully printed RIS prototype is shown in Fig. 13(b). The columns of the array are connected to the biasing lines with silver epoxy, as shown in Fig. 14. We can see that the $VO_2$ switches in the same column are serially connected.

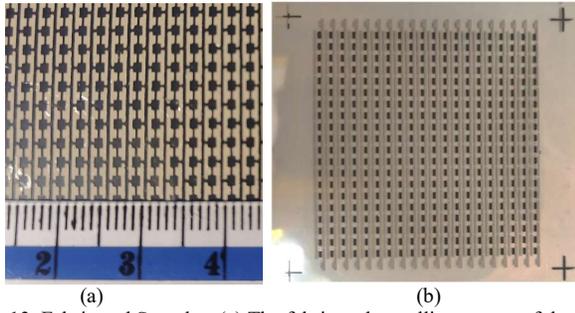

Fig. 13. Fabricated Samples. (a) The fabricated metallic patterns of the RIS. (b) The fabricated prototype of the RIS with VO2 switches.

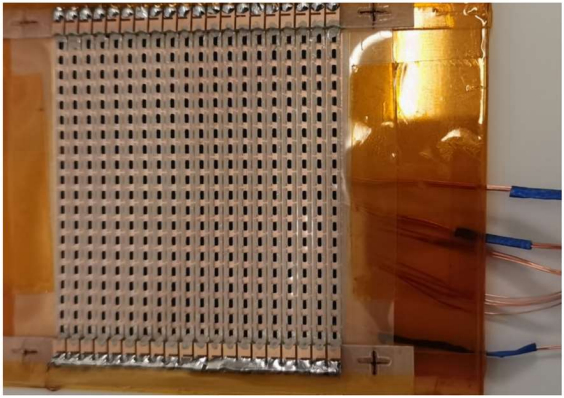

Fig. 14. The fabricated RIS array prototype with biasing

### B. Biasing Circuit Design and Fabrication

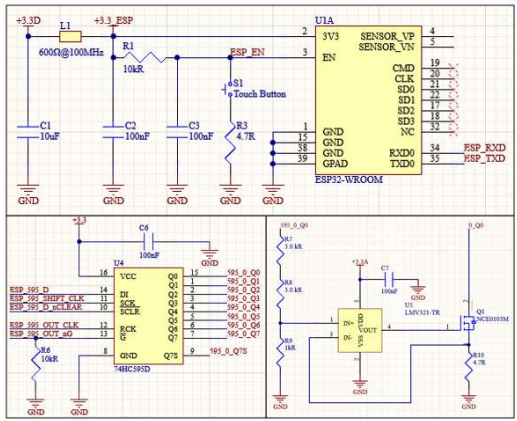

(a)

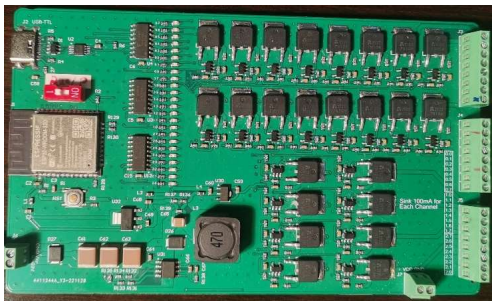

(b)

Fig. 15. The designed control circuits for the proposed RIS. (a) Core schematics and (b) PCB of the RIS controller.

To bias the proposed RIS array, we have designed and fabricated the RIS controller, as shown Fig. 15. In Fig. 15(a), ESP32 is the main controller, which reads commands from USB-TTL and drives the shift registers 74HC595 through a serial interface. The parallel outputs of the shift registers provide the reference voltage of the op-amp, and the op-amps control the sink current of the NMOS using the feedback loop. To drive the 20 serially connected unit cells, the controller requires 40V during the OFF to ON state transition, the $VO_2$ can remain in the ON state with 100 mA biasing current, which is defined with the 4.7-ohm current feedback resistor and the reference input voltage divider.

### C. RIS Near-field Measurement

Different from the reflect array antennas, RIS is expected to work in the near as well as far-field for both, the transmitter and the receiver [27 - 29]. In our experimental setup, two WR-34 waveguide standard gain horn antennas (PE9851B/SF-20) are used, which operate from 22 to 33 GHz with a nominal gain of 20 dBi. The Rx horn represents a user, while the Tx horn acts as the base station or another user in practical scenarios.

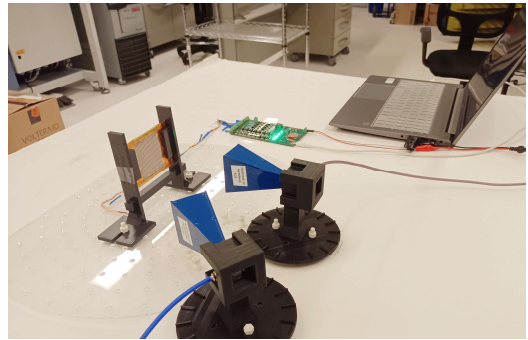

(a)

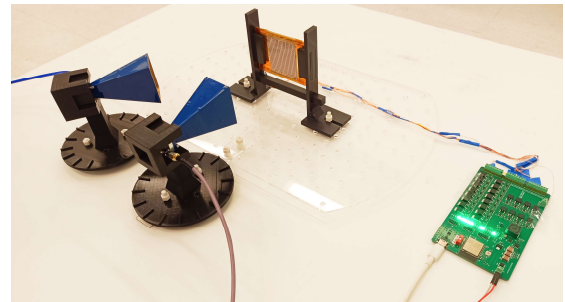

(b)

Fig. 16. Measurement setup for the characterization of the proposed fully-printed RIS design in the near field. (a) left view. (b) right view.

Since the designed RIS has an angular stability of up to 45°, the experiments have been conducted with an incidence angle of either 30° or 45°. Meanwhile, the receiver is placed at 0°, which is the desired reflection angle. It must be mentioned that these angles, in the measurements, are aligned with the horns' phase centers.

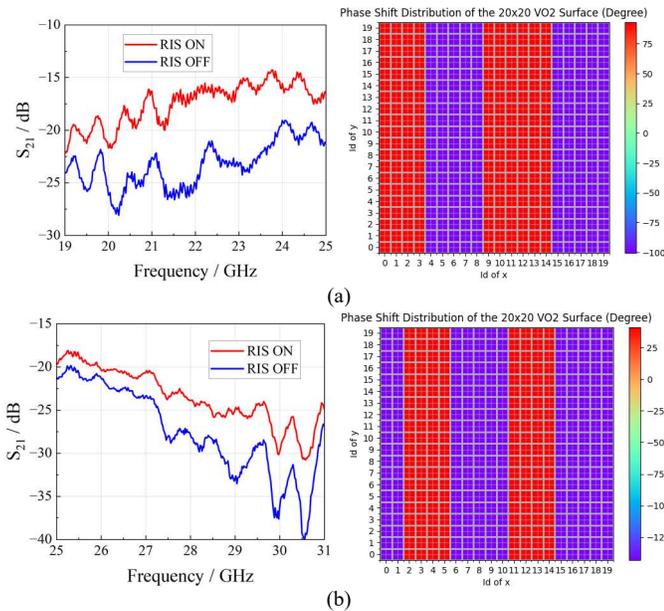
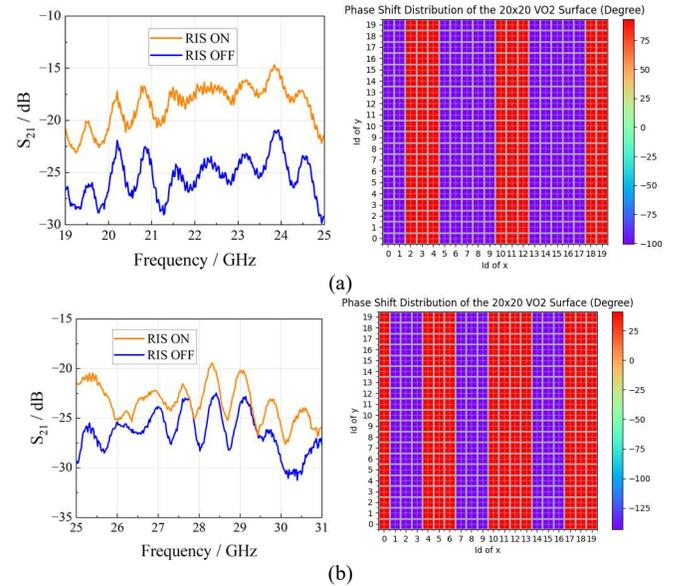

Fig. 17. Measured gain enhancement of the fabricated fully-printed RIS when turning ON the RIS pattern with 30° illumination compared with OFF state.

Fig. 18. Measured gain enhancement of the fabricated fully-printed RIS when turning ON the RIS pattern with 45° illumination compared with OFF state.

Fig. 16 depicts the measurement setup for the RIS operating in the near-field. The fabricated RIS prototype is assembled inside a 3D printed fixture, which has angle notations. The Tx and Rx horns are also positioned on 3D printed fixtures. Special care has been taken to ensure that the heights of the two horn antennas and the RIS are well aligned. The horn antennas are connected to Anritsu ME7828A vector network analyzer (VNA) to measure the $S_{21}$ parameter, which represents the received signal strength. The RIS pattern can be reconfigured by sending the commands to the control circuits through the PC. The ON/OFF states of the LEDs on the circuit board represent the states of the unit cells in a column of the designed RIS. To characterize and validate the fabricated fully-printed RIS prototype, the gain enhancement of RISs is a vital figure of merit, which can demonstrate how much signal can be focused in the desired reflection direction. Such evaluation has also been adopted in previous studies [10], [22], [30], [31].

The RIS gain enhancement can be assessed from the received power level of the Rx horn antenna. In the near-field experiment, the Tx horn antenna is placed at a distance of 20 cm from the RIS with and angle of incidence of 30°. The Rx horn antenna is also placed at a distance of 20 cm from the RIS but positioned at a refection angle of 0°. For this near-field measurement case, RIS receives a spherical wave from the Tx horn antenna. The measured results are shown in Fig. 17. Two different patterns and their respective gain enhancements are shown. Fig. 17(a) shows a higher gain enhancement at the lower frequency band (19 – 25 GHz), while Fig. 17(b) shows better performance at the higher frequency band (25 – 31 GHz). The measured results show that the highest gain enhancements in the lower and higher bands are 9.4 dB at 21.6 GHz and 9.5 dB at 30.5 GHz, respectively. Here, it must be mentioned that we just show the results for two patterns, as a proof of concept. In fact, for best performance, an optimum RIS pattern must be calculated for every frequency point. More sophisticated algorithms can be used optimized performance, however, that is beyond the scope of this work.

Considering the designed RIS has an angular stability of up to 45°, the same near-field measurement setup has been used for more experiments but with an angle of incidence of 45°. The measured results are illustrated in Fig. 18. A peak gain enhancement of 9.1 dB at 21.4 GHz and 8.0 dB at 25.1 GHz can be observed. As the incident angle increases to 45°, both the radar cross section (RCS) of the RIS as well as the reflection phase of the unit cell decreases. Thus, the measured results, with 45° incidence angle, show slightly lower gain enhancement as compared to the 30° incidence angle results. Nonetheless, the gain enhancement levels are still acceptable. It can be concluded that the fully printed RIS prototype can provide a gain enhancement of around 8 – 10 dB for a wide operational bandwidth. The gain enhancement values of the fully-printed RIS are somewhat lower than the previously reported PCB based designs [22], [30], [31], which is expected due to relatively larger conductor loss of the screen-printed metallic patterns, lower ON/OFF resistance ratio of the fully-printed $VO_2$ switches, 1-bit phase quantization, and the entire column based bias network. However, the performance is acceptable considering that this is achieved at a fraction of the cost of the PCB based designs with semiconductor switches. The performance can be further improved by enhancing the electrical properties of the $VO_2$ switches as well as the conductivity of the silver paste.

*D. RIS Far-field Measurement*

Since the RIS is expected to enhance the gain in the far field as well, far-field measurements have also been conducted. In this case plane wave illumination has been used. This is ensured by maintaining a distance of $2D^2/\lambda$ between the RIS and the horn antennas, where $D$ is the RIS aperture size, and $\lambda$ is the frequency of operation. Here, the horn antenna can be modeled as a point source radiated from its phase center [32]. For the proposed RIS design, the far-field distance is calculated to be 80 cm, thus the Tx-RIS and RIS-Rx are placed 80 cm away from each other, as shown in Fig. 19. In this measurement setup, the Tx horn is connected to a signal generator (Keysight VXG M9384B), while the Rx horn is connected to a spectrum analyzer (Keysight UXA N9040B). The gain enhancement of the RIS is proportional to the RCS in the far field, and the transmitted and received powers for the Tx and Rx horn antennas can be theoretically obtained from the radar range equation [32].

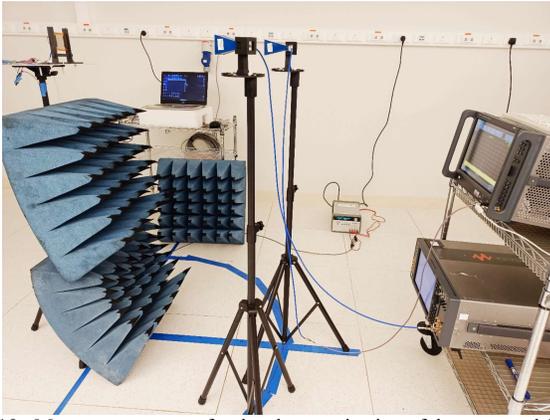

Fig.19 . Measurement setup for the characterization of the proposed fully-printed RIS design in the far field.

$$\frac{P_r}{P_t} = e_{cdt}e_{cdr}(1 - |\Gamma_t|^2)(1 - |\Gamma_r|^2)\sigma \frac{D_t D_r}{4\pi}$$
$$\times \left(\frac{\lambda}{4\pi R_1 R_2}\right)^2 |\hat{\rho}_w \cdot \hat{\rho}_r|^2 \qquad (2)$$

where $\sigma$ is the RCS of the RIS. Therefore, the received power level after reflection from the RIS can be obtained. Three tripods are deployed to ensure the Tx horn, RIS, and Rx horn are in the same horizontal plane. Some absorbers are attached to the tripod that supports the RIS to reduce the RCS from the metallic part of the tripod.

The received power levels for the ON and OFF states of the RIS, with 30° and 45° incidence angles and 0° reflection angle, are shown in Fig. 20 (a) and (b), respectively. Though, the measured results for only one frequency point (23.5 GHz) are shown here, other frequency points within the 5G mm-wave band demonstrate similar performance. From the measured results, it can be observed that the received power gets enhanced for both angles of incidence. The gain enhancements provided by the RIS for 30° and 45° incidence angles are 7.5 dB and 4.9 dB, respectively. The measured results, though lower than the near-field case, are still reasonable considering a lower spillover efficiency. The gain enhancement of 45° incidence angle case is lower than the 30° incidence angle case because of the lower effective receiving aperture size. Nevertheless, the measured results have proven that the proposed fully-printed RIS can provide gain enhancement in the far-field. Therefore, it can be concluded that the proposed fully-printed RIS based on VO$_2$ switches can provide decent gain enhancements in 5G mmWave band for both the near-field and far-field cases, and is thus is a promising candidate for low-cost and mass manufacturable mmWave RIS.

## VI. CONCLUSION

In conclusion, we, for the first time, proposed a fully VO$_2$-screen-printed wideband RIS from 23.5 GHz to 29.5 GHz. We have optimized the VO$_2$ ink preparation and printing process, including alignment and printing recipe, and have shown the decent performance of the screen-printed switch for mmWave RIS. We have combined the biasing and resonator of the RIS for a via-less design. Both the simulation and measurement results show that the developed RIS provides signal enhancement of 8 – 10 dB from 23.5 GHz to 29.5 GHz.

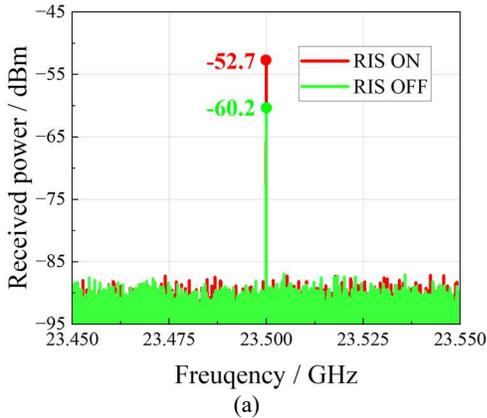

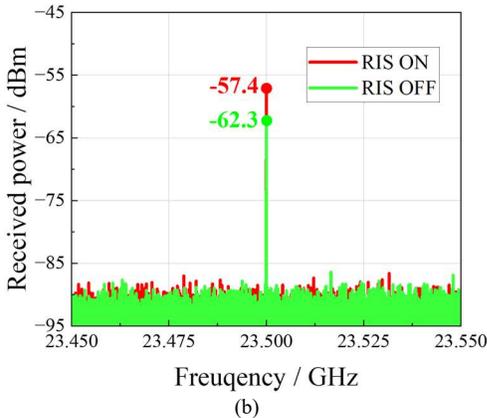

Fig. 20. Measured received power level of the Rx horn when the RIS at ON/OFF states at 23.5 GHz. (a) 30° illumination. (b) 45° illumination.